\documentclass[aps,prb,a4paper,preprint,showpacs]{revtex4}
\usepackage{amsmath}
\usepackage{amsfonts}
\usepackage{amssymb}
\usepackage{slashbox}
\usepackage{graphicx}
\usepackage{amsbsy}

\begin{document}
\title{Fano resonance in the nonadiabatic pumped shot noise of a time-dependent quantum well}
\author{Jiao-Hua Dai and Rui Zhu\renewcommand{\thefootnote}{*}\footnote{Corresponding author.
Electronic address:
rzhu@scut.edu.cn} }
\address{Department of Physics, South China University of Technology,
Guangzhou 510641, People's Republic of China }

\begin{abstract}

We use the Floquet scattering theory to study the correlation properties of the nonadiabatic pumped dc current and heat flow through a time-dependent quantum well. Electrons can transit through the quasibound state to the oscillator induced Floquet states leading to resonant tunneling effect. Virtual electron scattering processes can produce pumped heat flow, pumped shot noise and pumped heat flow noise, with presence of time and spatial reversal symmetry. When one of the Floquet levels matches the quasibound level there strikes a ``Fano" resonance.

\end{abstract}


\maketitle

\narrowtext

\section{Introduction}

Quantum pumping is a transport phenomenon originally proposed by Thouless\cite{ThoulessPRB1983} and first realized by Switkes \emph{et al}.\cite{SwitkesScience1999} It proposed that directed current can be induced by time-dependent modulation of external and internal parameters without bias in a quantum phase coherent nanoscale conductor. Theoretical and experimental research of quantum pumping has become a very important and active direction in mesoscopic physics. It is also significant in the field of quantum dynamic theory. By scale of the modulation frequency the quantum pump can be categorized into the adiabatic and nonadiabatic ones, with the former frequency much smaller than the characteristic tunneling times and vice the latter\cite{MoskaletsPRB2002, BrouwerPRB1998}. Adiabatic quantum pumping can be described by Berry phase of the scattering matrix accumulated during the cyclic modulation in the parameter space\cite{BrouwerPRB1998, XiaoRMP2010} and also by the energy quanta absorption and emission processes equivalent to the nearest sideband
approximation\cite{MoskaletsPRB2002}. Nonadiabatic quantum pumping can be described by the Floquet scattering scheme picturing quantities of interest in terms of sideband formation\cite{MoskaletsPRB2002Floquet, MoskaletsPRB2004}. The non-equilibrium Green's function\cite{BaigengWangPRB2002,BWangPRB2002,BWangPRB2003,ArracheaPRB2005,ArracheaPRB2007}, equation of motion\cite{AgarwalJPCM2007}, Galileo' transformation\cite{DasPLA2010}, and etc. also show physics of nonadiabatic quantum pumping from different views. Adiabatic and nonadiabatic quantum pumping has been investigated in various mesoscopic systems, such as nanowire\cite{DevillardPRB2008,XLQiPRB2009,SLZhuPRB2002,RZhuPRB2010,LebanonPRL2005}, mesoscopic rings\cite{CitroPRB2006}, quantum-dot structures\cite{KouwenhovenPRL1991,StrassPRL2005,ParkPRL2008,SplettstoesserPRB2008,
HattoriPRB2008,RomeoPRB2009,BrouwerPRB1998}, spin-orbit coupled conductors\cite{YCXiaoPLA2013}, magnetic tunnel junction\cite{RomeoEPJB2006}, graphene\cite{ZhuAPL2009,PradaPRB2009,TorresAPL2009,RochaPRB2010,WakkerPRB2010,TiwariAPL2010,
QZhangAPL2011,RZhuJPCM2011,SanJosePRB2011}, and superconductor junction with Majorana fermions\cite{AlosPalopPRB2014}, etc.

Current fluctuations are present in almost all kinds of
conductors including dynamical transport systems. The shot noise is the quantum contribution in the current fluctuation produced by the quantum coherence of charge carriers, which can give rich physical information in mesoscopic transport systems and is more significant in nanoscale quantum devices than in the traditional non-quantum devices\cite{BlanterPR2000}. Although intensive work has been done on the bias driven shot noise in various mesoscopic conductors\cite{BlanterPR2000,ZhuNova2011}, adiabatic pumped noise is also extensively investigated\cite{PolianskiPRB2002, DevillardPRB2008}, and the general scattering theory for nonadiabatic pumped shot noise\cite{MoskaletsPRB2004} is derived, the specific pumped shot noise properties in different quantum transport systems are less covered. They represent the underlying physics of different materials and devices, some of which is beyond conductance information.

Generally a transport approach covers the shot noise as well as the conductance.
The Floquet scattering matrix approach was developed for nonadiabatic noise properties as detailed by Moskalets \emph{et al}.\cite{MoskaletsPRB2004} General expressions for the pumped current, heat flow, and shot noise are derived for adiabatically and non-adiabatically driven quantum pumps. This approach stresses the existence of sidebands of electrons passing the time-dependent scatterer and these sidebands are connected to the currents and noise directly. Recently, Park and Ahn\cite{ParkPRL2008} derived an expression for the admittance and the current noise for a driven nanocapacitor in terms of the Floquet scattering matrix and obtained a non-equilibrium fluctuation dissipation relation. The scattering matrix renormalized by interaction has been used by Devillard \emph{et al}.\cite{DevillardPRB2008} to study the effect of weak electron-electron interaction on the noise. Under the geometric framework, there have been beautiful mathematical descriptions from the current to the noise\cite{DevillardPRB2008,PolianskiPRB2002}.

In this work, we focus on the non-adiabatic quantum pump driven by a single oscillating potential well, in which Fano resonance is predicted in the Floquet transmission spectrum when one of the Floquet levels matches the quasibound level of the static potential well\cite{WLiPRB1999}. In this case, the pumped current vanishes due to time and spatial reversal symmetry and the Fano resonance is unavailable in the current measurement. However, energy and information is transfused into the pump from exterior by instantaneous transport within a driving cycle. In the pumping process, virtual or temporary transmission within a cyclic period generates considerable noise. Supposing the resonance feature can be characterized in the correlation, we investigated its noise and heat flow properties.

\section{Model and numerical calculations}
 We consider a one-dimensional width-$L$ time-dependent potential-well sketched in Fig. 1. The time-dependent potential, which oscillates with frequency  $\omega $ and is located between $x = 0$ and $x = L$, has the form
\begin{equation}
U\left( {x,t} \right) = \left\{ \begin{array}{l}
 0,{\kern 90pt}  others, \\
 -U_0  + U_1 \cos \left( {\omega t} \right),{\kern 10pt}  0 < x < L . \\
 \end{array} \right.
\end{equation}
The time-dependent Hamiltonian of the electrons can be expressed as
\begin{equation}
H\left( t \right) =  - \frac{{\hbar ^2 }}{{2m^* }}\frac{{\partial ^2 }}{{\partial x^2 }} + U\left( {x,t} \right).
\end{equation}
$m^ *   = 0.067m_e $ is the effective mass of electrons and our discussion is based on single electron approximation and coherent tunneling.

Firstly, we consider the quasibound states within the one-dimensional static quantum well. When the electron is confined in the well with its energy $ E > - U_0 $, the wave function can be written as
\begin{equation}
\psi \left( x \right) = \left\{ {\begin{array}{*{20}{c}}
   {r{e^{\kappa x}},} \hfill & {x < 0 ,} \hfill  \\
   {a{e^{ikx}} + b{e^{ - ikx}},} \hfill & {0 < x < L ,} \hfill  \\
   {t{e^{ - \kappa x}},} \hfill & {x > L ,} \hfill  \\
\end{array}} \right.
\end{equation}
where $k={\sqrt {2 m^* (E+U_0)}}/ \hbar $ and $\kappa ={\sqrt {2 m^* (-E)}}/ \hbar $. Continuity equations of the wave function and its derivative at $x=0$ and $x=L$ are
\begin{equation}
\left\{ \begin{array}{l}
 r = a + b, \\
 \kappa r = ika - ikb, \\
 a{e^{ikL}} + b{e^{ - ikL}} = t{e^{ - \kappa L}}, \\
 ika{e^{ikL}} - ikb{e^{ - ikL}} =  - \kappa t{e^{ - \kappa L}}. \\
 \end{array} \right.
\end{equation}
Solvability of these equations gives rise to the secular equation
\begin{equation}
\left| {\begin{array}{*{20}{c}}
   1 & { - 1} & { - 1} & 0  \\
   \kappa  & { - ik} & {ik} & 0  \\
   0 & {{e^{ikL}}} & {{e^{ - ikL}}} & { - {e^{ - \kappa L}}}  \\
   0 & {ik{e^{ikL}}} & { - ik{e^{ - ikL}}} & {\kappa {e^{ - \kappa L}}}  \\
\end{array}} \right| = 0.
\end{equation}
Roots of $E$ for this equation are the quasibound state energies. It can be obtained numerically that the only quasibound state for the considered well configuration is at the energy\cite{WLiPRB1999} of $E_b = -0.17382 $ meV.

We use the Floquet scattering theory to investigate the quantum pumping properties of the oscillating quantum well\cite{WLiPRB1999}. Wave functions in the three scattering regimes can be written as
\begin{equation}
\begin{array}{l}
 \begin{array}{*{20}c}
   {\Psi _L \left( {x,t} \right) = \sum\limits_{n =  - \infty }^\infty  { e^{ - iE_n t/\hbar } \left(  {a_n^l e^{ik_n x}   + b_n^l e^{ - ik_n x}  } \right)} ,} & {x \le  0,}  \\
\end{array} \\
 \begin{array}{*{20}c}
   {\Psi _M \left( {x,t} \right) = \sum\limits_{n =  - \infty }^\infty  { e^{ - iE_n t/\hbar }  \sum\limits_{m =  - \infty }^\infty  {\left( {a_m e^{i\kappa _m x}  + b_m e^{ - i\kappa _m x} } \right)} J_{n - m} \left( {\frac{{U_1 }}{{\hbar \omega }}} \right) } ,} & { 0 \le x \le L,}  \\
\end{array} \\
 \begin{array}{*{20}c}
   {\Psi _R \left( {x,t} \right) = \sum\limits_{n =  - \infty }^\infty  {  e^{ - iE_n t/\hbar }  \left( {a_n^r e^{ - ik_n x}  + b_n^r e^{ik_n x}  } \right)} ,} & {x \ge L.}  \\
\end{array} \\
 \end{array}
 \label{WaveFunction}
\end{equation}
In the left and right free regions, the incident and outgoing electron waves consist of infinite number of sidebands, as shown in Fig. 1. $E_n  = E_F  + n\hbar \omega $ is the eigenenergy of the $n$-th order Floquet state with the Fermi energy $E_F $ and $k_n  = \sqrt {2m^ *  E_n  }/ \hbar $ is the corresponding wave vector. $n$ are integers varying from $ - \infty $ to $ + \infty $ in an ideal exactness. While $E_n  < 0$, $k_n $ is imaginary meaning an evanescent mode, the current for this channel vanishes. $a_n^{l/r} $ and $b_n^{l/r} $ are the probability amplitudes corresponding to those flowing out of and flowing into the left/right leads, respectively. Here we did not use flux normalization for algebra simplicity of continuity conditions and its justification is elaborated in the appendix. The Floquet scattering matrix can be constructed by relations between $a_n^{l/r} $ and $b_n^{l/r} $. For example, if we set $a_0^l  = 1 $, $b_n^l $ corresponds to the reflection amplitude from the left lead to the left lead in the  $n$-order Floquet channel, and so on. $\kappa _m  =  \sqrt {2m^ *  \left( {E_F  + m\hbar \omega  + U_0 } \right)} /\hbar $ is wavevector in the middle oscillating quantum well region. $J_n $ is the first kind of Bessel function deriving from $\exp \left( { - ix\sin \beta } \right) = \sum\nolimits_{n =  - \infty }^\infty  {J_n \left( x \right)e^{in\beta } } $, which only exists in the oscillating region. $a_m $ and $b_m $ are the wave function amplitudes in the oscillating region and present only in the continuity equation solving processes.

By solving equations of the boundary conditions at interfaces
\begin{equation}
\Psi _L \left( {0,t} \right) = \Psi _M \left( {0,t} \right),\begin{array}{*{20}c}
   {} & {\Psi _M \left( {L,t} \right) = \Psi _R \left( {L,t} \right),}  \\
\end{array}
\end{equation}
and
\begin{equation}
\frac{{\partial \Psi _L \left( {0,t} \right)}}{{\partial x}} = \frac{{\partial \Psi _M \left( {0,t} \right)}}{{\partial x}},{\kern 20pt}  \frac{{\partial \Psi _M \left( {L,t} \right)}}{{\partial x}} = \frac{{\partial \Psi _R \left( {L,t} \right)}}{{\partial x}}{\kern 3pt} ,
\end{equation}
which must hold for all time, the Floquet scattering matrix without flux normalization can be obtained by matrix algebra. It connects different Floquet modes as
\begin{equation}
\left( {\begin{array}{*{20}{c}}
   {b_n^l}  \\
   {b_n^r}  \\
\end{array}} \right) = \sum\limits_{m =  - \infty }^{ + \infty } {\left( {\begin{array}{*{20}{c}}
   {{r_{nm}}} & {t{'_{nm}}}  \\
   {{t_{nm}}} & {r{'_{nm}}}  \\
\end{array}} \right)\left( {\begin{array}{*{20}{c}}
   {a_m^l}  \\
   {a_m^r}  \\
\end{array}} \right)}  = \sum\limits_{m =  - \infty }^{ + \infty } {\left( {\begin{array}{*{20}{c}}
   {{S_{LLnm}}} & {{S_{LRnm}}}  \\
   {{S_{RLnm}}} & {{S_{RRnm}}}  \\
\end{array}} \right)\left( {\begin{array}{*{20}{c}}
   {a_m^l}  \\
   {a_m^r}  \\
\end{array}} \right)} ,
\label{eq6}
\end{equation}
with $a^{l/r} $ and $b^{l/r} $ column vectors made up of $a_n^{l/r} $ and $b_n^{l/r} $ of all  $n'$s. $t_{nm} / t'_{nm}$ and $r_{nm} / r'_{nm}$ are the transmission and reflection amplitudes incoming from the $m$-th Floquet channel and going into the $n$-th Floquet channel. Considering the real current flux, the Floquet scattering matrix elements are
\begin{equation}
{s_{\alpha \beta }}\left( {{E_n},{E_m}} \right) = \sqrt {\frac{{{\mathop{\rm Re}\nolimits} \left( {{k_n}} \right)}}{{{\mathop{\rm Re}\nolimits} \left( {{k_m}} \right)}}} {S_{\alpha \beta nm}}.
\label{FluxNormalizedScatteringMatrix}
\end{equation}
Transmission of evanescent modes would vanish as only the real part of the outgoing wave vectors is considered.
From the scattering matrix, the total transmission probability is defined as
\begin{equation}
T = {\sum\limits_{m,n =  - \infty }^\infty  {\frac{{{\mathop{\rm Re}\nolimits} \left( {k_n } \right)}}{{{\mathop{\rm Re}\nolimits} \left( {k_m } \right)}}\left| {t_{nm} } \right|^2 } }   .
\end{equation}

We are interested in the element $s_{\alpha \beta } \left( {E_n ,E} \right)$ of the Floquet scattering matrix given in Eq. (\ref{FluxNormalizedScatteringMatrix}). It measures the scattering amplitude of the electron incident through lead $\beta$ with energy  $E$ and leaving through lead $\alpha $ with energy $E_n $. By interacting with the oscillating potential the electron absorbs or loses energy quanta of $n\hbar \omega $, with its final energy $E_n  = E \pm n \hbar \omega $.

For the non-adiabatic quantum pump, the Floquet scattering matrix is sensitive to the spatial symmetry of the potential. If the system is present of perturbations broken the spatial symmetry or time-reversal symmetry it can pump a dc current. With only one oscillating potential, both the spatial symmetry and time-reversal symmetry is present in the device, thus no pumped current exists with
\begin{equation}
I_L=-I_R=0.
\end{equation}

The heat flow is carried by the non-equilibrium particles, which occurs in the process of scattering and the direction of heat flow is defined as from the oscillating potential to the reservoirs as
\begin{equation}
I_\alpha ^H  = \frac{1}{h}\int_0^\infty  {dE {\sum\limits_{n, \beta } {\left( {E_n  - \mu } \right)} } } \left| {s_{\alpha \beta } \left( {E_n ,E} \right)} \right|^2 \left[ {f_0 \left( E \right) - f_0 \left( {E_n } \right)} \right].
\label{eq9}
\end{equation}
Here $f_0 \left( E \right)$ is equilibrium Fermi distribution function. $\mu$ is the chemical potential, which is the same in all reservoirs at zero bias. We also assume all the reservoirs have the same temperature.

The problem of current noise is closely connected with the matrix elements of $s_{\alpha \beta } \left( {E_n ,E} \right)$. For a phase-coherent conductor the noise is sensitive to the quantum-mechanical interference effects. We can describe the correlation function of the current as\cite{BlanterPR2000, ButtikerPRB1992}
\begin{equation}
{S}_{\alpha \beta } \left( {t_1 ,t_2 } \right) = \frac{1}{2}\left\langle {\hat I_\alpha  \left( {t_1 } \right)\hat I_\beta  \left( {t_2 } \right) + \hat I_\beta  \left( {t_2 } \right)\hat I_\alpha  \left( {t_1 } \right)} \right\rangle {\kern 1pt} {\kern 1pt} ,
\label{eq1}
\end{equation}
where $\Delta \hat I = \hat I - \left\langle {\hat I} \right\rangle $, and $\hat I_\alpha ^{} \left( t \right)$ is the quantum-mechanical current operator in the lead $\alpha $, which can be expressed as\cite{BlanterPR2000}
\begin{equation}
\hat I_\alpha  \left( t \right) = \frac{e}{h}\int {dEdE'\left[ {\hat b_\alpha ^\dag  \left( E \right)\hat b_\alpha  \left( {E'} \right) - \hat a_\alpha ^\dag  \left( E \right)\hat a_\alpha  \left( {E'} \right)} \right]} e^{i\left( {E - E'} \right)t/\hbar },
\label{eq11}
\end{equation}
with $\hat a_\alpha  \left( E \right)$ and $\hat b_\alpha  \left( E \right)$ annihilation operators of the incident and outgoing electrons to the driven potential and
\begin{equation}
{\hat b_\alpha }\left( E \right) = \sum\limits_{n, \beta} {{s_{\alpha \beta }}\left( {E,{E_n}} \right){{\hat a}_\beta }\left( {{E_n}} \right)} .
\label{bsa}
\end{equation}

From Eqs. (\ref{eq1}) to (\ref{bsa}) the pumped shot noise and pumped heat flow noise can be expressed in terms of the Floquet scattering matrix as\cite{MoskaletsPRB2004}
\begin{equation}
S_{\alpha \beta }  = \frac{{e^2 }}{h}\int_0^\infty dE {\sum\limits_{\gamma ,\delta } { { {\sum\limits_{m,n,p =  - \infty }^\infty  {{\rm M}_{\alpha \beta \gamma \delta }(E, E_m, E_n, E_p) \left[ {f_0 \left( {E_n } \right) - f_0 \left( {E_m } \right)} \right]^2 } } } } } ,
\label{eq2}
\end{equation}
\begin{equation}
S_{\alpha \beta }^H  = \frac{1}{h}\int_0^\infty  {dE\sum\limits_{\gamma ,\delta } { { {\sum\limits_{m,n,p =  - \infty }^\infty  {{\rm M}_{\alpha \beta \gamma \delta }(E, E_m, E_n, E_p) \left( {E - \mu } \right)\left( {E_p  - \mu } \right)} } } } } \left[ {f_0 \left( {E_n } \right) - f_0 \left( {E_m } \right)} \right]^2 ,
\label{eq3}
\end{equation}
with
\begin{equation}
{\rm M}_{\alpha \beta \gamma \delta }(E, E_m, E_n, E_p) {\rm{ = }}s_{\alpha \gamma }^ *  \left( {E,E_n } \right){\kern 1pt} s_{\alpha \delta }^{} \left( {E,E_m } \right)s_{\beta \delta }^ *  \left( {E_p ,E_m } \right)s_{\beta \gamma }^{} \left( {E_p ,E_n } \right)
\end{equation}
describing the quantum-mechanical exchange during scattering of electrons with energy $E_n ,{\kern 1pt} {\kern 1pt} {\kern 1pt} E_m $ incident from leads $\gamma ,{\kern 1pt} {\kern 1pt} {\kern 1pt} \delta $ and outgoing to the leads $\alpha ,{\kern 1pt} {\kern 1pt} {\kern 1pt} \beta $ with energy $E,{\kern 1pt} {\kern 1pt} {\kern 1pt} E_p $, respectively.

Current flux conservation secures that $S_{LL}=S_{RR}=-S_{LR}=-S_{RL}$ and $I_L^H=I_R^H$. We consider one of the four and label $S_{LL}$ as $S_I$, $S^H _{LL}$ as $S_H$, and $I_L^H=H$. To magnify the resonance spectrum, we also considered the derivatives of the noise and heat flow over the Fermi energy with
\begin{equation}
\begin{array}{*{20}{c}}
   {S_{I/H}^d = \frac{{dS_{I/H}^d}}{{d{E_F}}},} & {{H^d} = \frac{{dH}}{{d{E_F}}}}  \\
\end{array}.
\end{equation}

\section{Numerical results and discussion}

In this paper we have adopted the Floquet scattering matrix approach to investigate the pumped effect of phase coherent mesoscopic systems of noninteracting electrons. The Floquet scattering matrix describes existence of sidebands of electrons entering and exiting the pump. The nonequilibrium electrons generated by the pump carry heat from the oscillating potential to the reservoirs and transfer charge between the two reservoirs. Only the first sidebands $\left( {n =  \pm 1} \right)$ are excited if the oscillating amplitude is small.

The total transmission probability $T = \sum\nolimits_{n = 0}^5 {{{\left| {{t_{0n}}} \right|}^2}} $ as a function of the incident energy is shown in Fig. 2. In all of the numerical consideration $\hbar \omega$ is set to be 1 meV. We take into account different Floquet sidebands both above and below the incident energy, with $n = 0, \pm 1,..., \pm N$. $N=5$ cutoff is used, with its precision satisfactory for the small driving amplitude. In the quantum well there exists a quasibound state, when the first order Floquet sideband overlaps with the quasibound level, a ``Fano" resonance occurs (also confer Fig. 1), which was discovered in Ref. \onlinecite{WLiPRB1999}. Due to time and spatial reversal symmetry, no charge current is generated by a single oscillating quantum well. It is known that when pumped charge current is zero, the pumped shot noise can be considerably large due to virtual transmission processes\cite{RZhuPRB2010, MoskaletsPRB2004}. We suppose that the nonequilibrium transmission properties can be recorded in the shot noise spectrum and the ``Fano" resonance can thus be observed.

We calculated the pumped current noise, heat noise and heat flow driven by the nonadiabatic oscillating quantum well using the Floquet scattering scheme with Eqs. (\ref{eq9}), (\ref{eq2}), and (\ref{eq3}). Their variation as a function of the Fermi energy was depicted in Fig. 3. The pumped current noise, heat noise and heat flow increases with the Fermi energy when more energy channels contribute to the transport for larger Fermi energies. For a small driving amplitude in our case, most of the transmission comes from the original incident level and the two first order Floquet sidebands ($n=0,\pm 1$). When these bands completely go out of the quantum well with $E_F \approx \hbar \omega$, a decrease occurs in the noise spectrum. Noise is an effect of correlation, concrete or virtual. When all the active Floquet bands are out of the quantum well, the transmitting electron ``sees" no structure in the conductor therefore ballistic transport governs giving rise to the shot noise decrease. An inflection could be found at the Fermi energy $E_F \approx 0.826$ meV corresponding to the ``Fano" resonance in the total transmission. The magnification of the inflection point is shown in the insets. Since the transport properties are an accumulating effect of all energy channels, contribution of a single energy channel of the resonance is limited. To magnify the ``Fano" resonance of the nonadiabatic quantum pump, we calculated the differentials of the pumped charge and heat noise to the fermi energy and dramatic resonance pattern could be found.

Differentials of the pumped current noise, heat noise, and heat flow as a function of the Fermi energy are shown in Fig. 4. These curves have a sharp dip followed by a peak at $E_F  \approx 0.826 $ meV, demonstrating an asymmetric ``Fano" resonance, which ensures that there exists a quasibound state (the energy of the quasibound state is $E_B  \approx  - 0.174 $ meV) in the deep quantum well\cite{WLiPRB1999}. Electrons in the propagating states can emit photons and drop into the quasibound state and bounce back before exiting the well, thus contributing to the transport (see Fig. 1). The heat flow also shows a sharp peak at the resonance Fermi energy. The pumped noise properties can be interpreted as follows. At the resonant Fermi energy, transport process and the electron-electron correlation achieve maximal strength. The nonequilibrium electrons created by the oscillating scatterer move in different directions carried the heat flow into the electron reservoirs of the two sides. The differential shot noise demonstrates peaks corresponding to the ``Fano" resonance, as a result of the virtual motion of nonequilibrium electrons.

\section{Conclusions}

We considered the noise properties of a nonadiabatic quantum pump driven by an oscillating potential well. Due to time and space reversal symmetry no dc charge current can be produced. Due to virtual transmission of the electrons, heat current is not zero with its direction from the conductor into the leads at both reservoirs. To experimentally observe the ``Fano" resonance found in the transmission\cite{WLiPRB1999}, we investigated the heat current and shot noise of the charge and heat current. Sharp ``Fano"-shape resonance was found. The differential current noise, heat noise, and heat flow demonstrate peak structure from the interaction of electrons with the oscillating potential when one of the Floquet sideband matches the quasibound state. Electrons in incident channel can drop into the quasibound state by emitting photons. Similarly, electron in the bound state can also absorb photons and bounce back into the Floquet channels. Thus a ``Fano" resonance occurred. The resonance position of the Fermi energy is then governed by the energy of the static quasibound state.

\section{Acknowledgements}

The authors acknowledge enlightening discussions with professor Wen-Ji Deng. This project was supported by the National Natural Science
Foundation of China (No. 11004063) and the Fundamental Research
Funds for the Central Universities, SCUT (No. 2014ZG0044).

\section{Appendix. Wave Function Normalization and Floquet Scattering Matrix Unitarity }

The current operator in the left lead (far from the sample) can be expressed in a standard way\cite{BlanterPR2000},
\begin{equation}
{\hat I_L}\left( {x,t} \right) = \frac{{\hbar e}}{{2i{m^*}}}\int {d{{\bf{r}}_ \bot }\left[ {\hat \psi _L^\dag \left( {{\bf{r}},t} \right)\frac{\partial }{{\partial x}}{{\hat \psi }_L}\left( {{\bf{r}},t} \right) - \left( {\frac{\partial }{{\partial x}}\hat \psi _L^\dag \left( {{\bf{r}},t} \right)} \right){{\hat \psi }_L}\left( {{\bf{r}},t} \right)} \right]} ,
\label{ACurrent}
\end{equation}
where the field operator $\hat \psi$ is defined as
\begin{equation}
{\hat \psi _L}\left( {\textbf{r},t} \right) = \sum\limits_{n =  - \infty }^{ + \infty } {\int {d{E_n}} {e^{{{ - i{E_n}t} \mathord{\left/
 {\vphantom {{ - i{E_n}t} \hbar }} \right.
 \kern-\nulldelimiterspace} \hbar }}}\frac{{{\chi _L}\left( {{\textbf{r}_ \bot }} \right)}}{{{{\left[ {2\pi \hbar {v_L}\left( {{E_n}} \right)} \right]}^{{1 \mathord{\left/
 {\vphantom {1 2}} \right.
 \kern-\nulldelimiterspace} 2}}}}}\left[ {{{\hat a}_L}\left( {{E_n}} \right){e^{i{k_{n}}x}} + {{\hat b}_L}\left( {{E_n}} \right){e^{ - i{k_{n}}x}}} \right]} .
 \label{AFieldOperator}
\end{equation}
To avoid tediousness we consider single transverse channel and multiple energy channels, the latter of which is necessary for the nonadiabatic dynamic process. $E_n$ and $k_n$ are defined identical to the  main text. It can be seen that the field operator is naturally flux normalized with ${\sqrt {v_L (E_n)}}$ in the denominator.

Substituting Eq. (\ref{AFieldOperator}) into Eq. (\ref{ACurrent}), and using the relations,
\begin{equation}
{v_L}\left( {{E_n}} \right) = \frac{{\hbar {k_{Ln}}}}{{{m^*}}},
\end{equation}
and
\begin{equation}
\int {\chi _L^*\left( {{{\bf{r}}_ \bot }} \right){\chi _L}\left( {{{\bf{r}}_ \bot }} \right)d{{\bf{r}}_ \bot }}  = 1,
\end{equation}
 we could have
\begin{equation}
\begin{array}{l}
 {{\hat I}_L}\left( {x,t} \right) = \frac{e}{{4\pi \hbar }}\sum\limits_{m,n =  - \infty }^{ + \infty } {\int {d{E_n}d{E_m}{e^{{{i\left( {{E_n} - {E_m}} \right)t} \mathord{\left/
 {\vphantom {{i\left( {{E_n} - {E_m}} \right)t} \hbar }} \right.
 \kern-\nulldelimiterspace} \hbar }}}\left\{ {\left[ {{{\left( {\frac{{{k_{Lm}}}}{{{k_{Ln}}}}} \right)}^{{1 \mathord{\left/
 {\vphantom {1 2}} \right.
 \kern-\nulldelimiterspace} 2}}} + {{\left( {\frac{{{k_{Ln}}}}{{{k_{Lm}}}}} \right)}^{{1 \mathord{\left/
 {\vphantom {1 2}} \right.
 \kern-\nulldelimiterspace} 2}}}} \right]} \right.} }  \\
  \times \left[ {{e^{i\left( {{k_{Lm}} - {k_{Ln}}} \right)x}}\hat a_L^\dag \left( {{E_n}} \right){{\hat a}_L}\left( {{E_m}} \right) - {e^{ - i\left( {{k_{Lm}} - {k_{Ln}}} \right)x}}\hat b_L^\dag \left( {{E_n}} \right){{\hat b}_L}\left( {{E_m}} \right)} \right] \\
 \left. { - \left[ {{{\left( {\frac{{{k_{Lm}}}}{{{k_{Ln}}}}} \right)}^{{1 \mathord{\left/
 {\vphantom {1 2}} \right.
 \kern-\nulldelimiterspace} 2}}} - {{\left( {\frac{{{k_{Ln}}}}{{{k_{Lm}}}}} \right)}^{{1 \mathord{\left/
 {\vphantom {1 2}} \right.
 \kern-\nulldelimiterspace} 2}}}} \right]\left[ {{e^{ - i\left( {{k_{Lm}} + {k_{Ln}}} \right)x}}\hat a_L^\dag \left( {{E_n}} \right){{\hat b}_L}\left( {{E_m}} \right) - {e^{i\left( {{k_{Lm}} + {k_{Ln}}} \right)x}}\hat b_L^\dag \left( {{E_n}} \right){{\hat a}_L}\left( {{E_m}} \right)} \right]} \right\} . \\
 \end{array}
\end{equation}

We consider the dc current driven by nonadiabatic periodic parameter variation. The time-averaged current can be calculated in an arbitrary pumping cycle as
\begin{equation}
{I_L} = \frac{\omega }{{2\pi }}\int_0^{{{2\pi } \mathord{\left/
 {\vphantom {{2\pi } \omega }} \right.
 \kern-\nulldelimiterspace} \omega }} {\left\langle {{{\hat I}_L}\left( {x,t} \right)} \right\rangle dt} ,
\end{equation}
where $\left\langle {\hspace{1mm}} \right\rangle $ means the quantum and statistical average of the current operator. With the integral
\begin{equation}
\int_0^{{{2\pi } \mathord{\left/
 {\vphantom {{2\pi } \omega }} \right.
 \kern-\nulldelimiterspace} \omega }} {{e^{{{i\left( {{E_n} - {E_m}} \right)t} \mathord{\left/
 {\vphantom {{i\left( {{E_n} - {E_m}} \right)t} \hbar }} \right.
 \kern-\nulldelimiterspace} \hbar }}}dt}  = \int_0^{{{2\pi } \mathord{\left/
 {\vphantom {{2\pi } \omega }} \right.
 \kern-\nulldelimiterspace} \omega }} {{e^{i\left( {n - m} \right)\omega t}}dt}  = \frac{{2\pi }}{\omega }\delta \left( {n - m} \right),
\end{equation}
we could obtain
\begin{equation}
{I_L} = \frac{e}{{2\pi \hbar }}\sum\limits_{n =  - \infty }^{ + \infty } {\int {d{E_n}\left[ {\left\langle {\hat a_L^\dag \left( {{E_n}} \right){{\hat a}_L}\left( {{E_n}} \right)} \right\rangle  - \left\langle {\hat b_L^\dag \left( {{E_n}} \right){{\hat b}_L}\left( {{E_n}} \right)} \right\rangle } \right]} } ,
\label{ACurrentWithn}
\end{equation}
Here it should be noted that the Floquet channels occur only in the scattering process and the original and final states are energy conserved with the energy of $E$. Then, we have
\begin{equation}
{I_L} = \frac{e}{{2\pi \hbar }}\int {dE\left[ {\left\langle {\hat a_L^\dag \left( E \right){{\hat a}_L}\left( E \right)} \right\rangle  - \left\langle {\hat b_L^\dag \left( E \right){{\hat b}_L}\left( E \right)} \right\rangle } \right]} .
\label{AAveragedCurrent}
\end{equation}
 The outgoing operators can be expressed in terms of the incoming ones by the Floquet scattering matrix
\begin{equation}
{{\hat b}_L}\left( {{E_n}} \right) = \sum\limits_{\beta ,p} {{{ s}_{L\beta }}\left( {{E_n},{E_p}} \right){{\hat a}_\beta }\left( {{E_p}} \right)} .
\label{ADefineSMatrix}
\end{equation}
For reservoirs at equilibrium, we have
\begin{equation}
\left\langle {\hat a_{\alpha }^\dag \left( {{E_n}} \right){{\hat a}_{\beta}}\left( {{E_m}} \right)} \right\rangle  = {f_{\alpha}}\left( E_n \right) {\delta _{\alpha , \beta}} {\delta (E_n-E_m)} .
\label{AFermiDistribution}
\end{equation}
Substituting Eqs. (\ref{ADefineSMatrix}) and (\ref{AFermiDistribution}) into the averaged current (\ref{AAveragedCurrent}), we can obtain
\begin{equation}
{I_L} = \frac{e}{{2\pi \hbar }}\int {dE\left[ {{f_L}\left( E \right) - \sum\limits_{\beta ,n} {{{\left| {{{ s}_{L\beta }}\left( {E,{E_n}} \right)} \right|}^2}{f_\beta }\left( {{E_n}} \right)} } \right]} ,
\end{equation}
which reproduced the result of Ref. \onlinecite{MoskaletsPRB2002Floquet}.

Unitarity of the Floquet scattering matrix naturally follows from the current conservation. The current flowing from the left lead is identical to that flowing into the right lead $ {I_L} =  - {I_R} $ (Positive direction of the current is defined as flowing from the reservoir to the scatterer). From Eq. (\ref{ACurrentWithn}), we have
\begin{equation}
\sum\limits_n {\left[ {\hat a_L^\dag \left( {{E_n}} \right){{\hat a}_L}\left( {{E_n}} \right) - \hat b_L^\dag \left( {{E_n}} \right){{\hat b}_L}\left( {{E_n}} \right)} \right]}  = \sum\limits_n {\left[ {\hat b_R^\dag \left( {{E_n}} \right){{\hat b}_R}\left( {{E_n}} \right) - \hat a_R^\dag \left( {{E_n}} \right){{\hat a}_R}\left( {{E_n}} \right)} \right]} .
\end{equation}
By Eq. (\ref{ADefineSMatrix}), it follows that
\begin{equation}
\sum\limits_{\alpha n} {\hat a_\alpha ^\dag \left( {{E_n}} \right){{\hat a}_\alpha }\left( {{E_n}} \right)}  = \sum\limits_{\alpha \beta \gamma pl} {\hat a_\alpha ^\dag \left( {{E_p}} \right)s_{\gamma \alpha }^*\left( {{E_n},{E_p}} \right){s_{\gamma \beta }}\left( {{E_n},{E_l}} \right){{\hat a}_\beta }\left( {{E_l}} \right)} ,
\end{equation}
which could be written into the matrix form as
\begin{equation}
{{\bf{\hat a}}^\dag }{\bf{\hat a}} = {{\bf{\hat a}}^\dag }{{\bf{\hat s}}^\dag }{\bf{\hat s\hat a}},
\label{ASMatrixUnitarity}
\end{equation}
with corresponding elements
\begin{equation}
{\left( {{\bf{\hat s}}} \right)_{\alpha \beta mn}} = {s_{\alpha \beta }}\left( {{E_m},{E_n}} \right),
\end{equation}
and
\begin{equation}
{\left( {{\bf{\hat a}}} \right)_{\alpha n}} = {\hat a_\alpha }\left( {{E_n}} \right).
\end{equation}
Directly from Eq. (\ref{ASMatrixUnitarity}) follows the unitarity relation
\begin{equation}
{{{\bf{\hat s}}}^\dag }{\bf{\hat s}} = {\bf{\hat I}}.
\end{equation}
All elements of the Floquet scattering matrix are only well defined for incident and outgoing propagating modes.

In deducing the Floquet scattering matrix from continuity conditions, it is more convenient to use the wave function without flux normalization as in Eq. (\ref{WaveFunction}). By considering the real current flux, we do the transform of Eq. (\ref{FluxNormalizedScatteringMatrix}) to obtain the Floquet scattering matrix. We could reproduce previous derivations by defining
\begin{equation}
\begin{array}{*{20}{c}}
   {{{\hat A}_\alpha }\left( {{E_n}} \right) = \frac{{{{\hat a}_\alpha }\left( {{E_n}} \right)}}{{\sqrt {{k_n}} }},} & {{{\hat B}_\alpha }\left( {{E_n}} \right) = \frac{{{{\hat b}_\alpha }\left( {{E_n}} \right)}}{{\sqrt {{k_n}} }},}  \\
\end{array},
\end{equation}
and
\begin{equation}
{\hat B_\alpha }\left( {{E_n}} \right) = \sum\limits_{\beta ,m} {{S_{\alpha \beta nm}}{{\hat A}_\beta }\left( {{E_m}} \right)} ,
\end{equation}
with $S_{\alpha \beta nm}$ defined in Eq. (\ref{eq6}) and directly obtainable from continuity relations.

\clearpage

\begin{figure}[h]
\includegraphics[height=7cm, width=10cm]{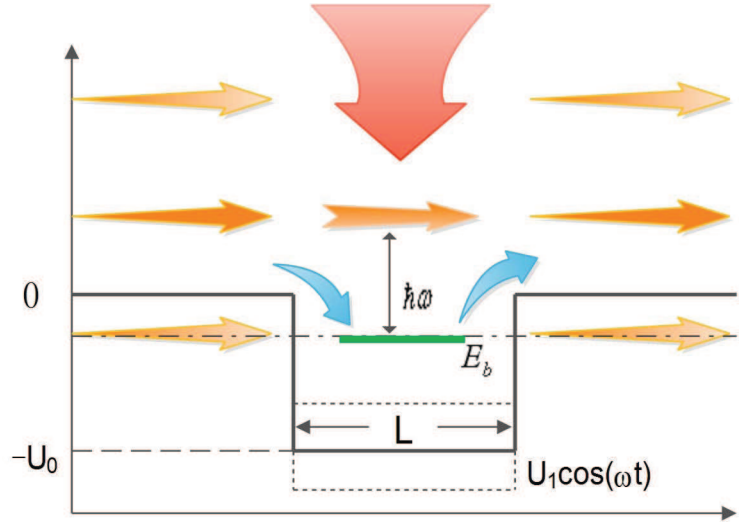}
\caption{Profile of the quantum pump formed by an oscillating quantum well. Two adjacent Floquet states have energy spacing of $\hbar \omega $. There is a quasibound state in the potential well with the binding energy $E_b$. Energy is infused into the system by ac modulation of the potential well. Fano resonance occurs when one of the Floquet sideband overlaps with the quasibound state. Equilibrium well depth is $U_0$ and its variation in time has the form of ${U_1}\cos \left( {\omega t} \right)$.}
\end{figure}

\clearpage

\begin{figure}[h]
\includegraphics[height=10cm, width=12cm]{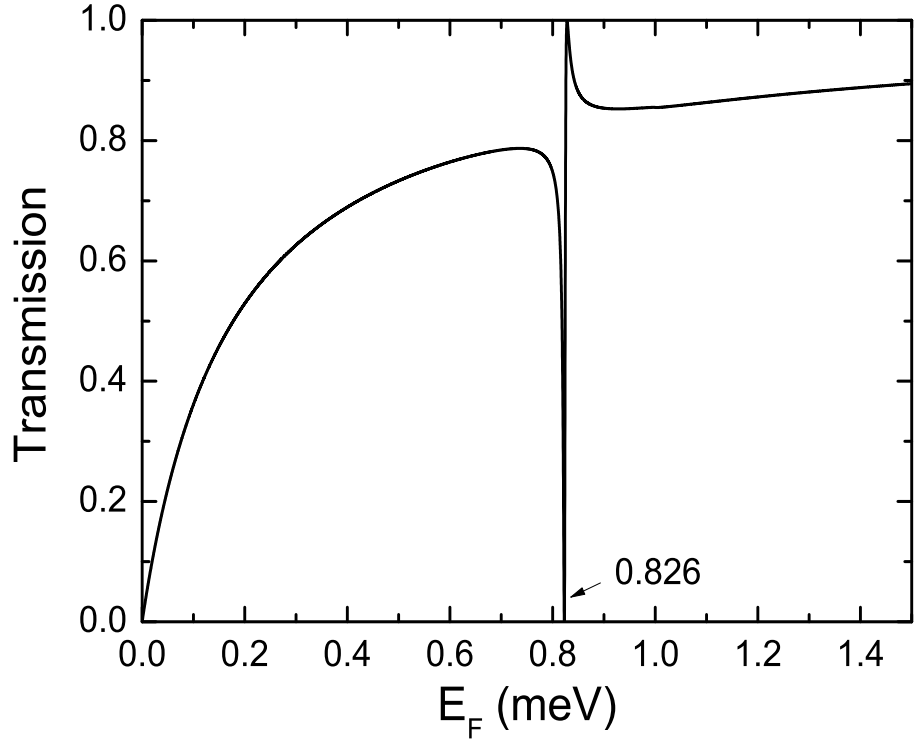}
\caption{Total transmission probability $T = \sum\nolimits_{n = 0}^5 {{{\left| {{t_{0n}}} \right|}^2}} $
 as a function of the incident energy\cite{WLiPRB1999}. Driving amplitude ${U_1} = 5$ meV, static well depth $U_0 = 20$ meV, well width $L=10$ \AA, and energy quanta of the driving frequency $\hbar \omega  = 1$ meV. A resonance occurs at $E_F \approx 0.826$ meV.
 }
\end{figure}

\begin{figure}[h]
\includegraphics[height=10cm, width=11cm]{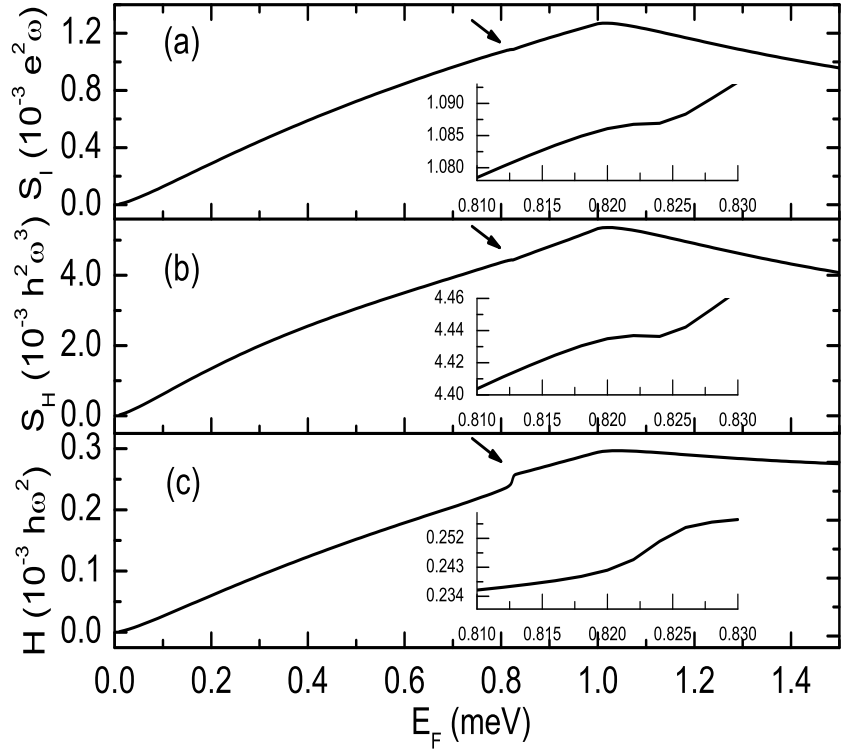}
\caption{(a) Current shot noise $S_I$, (b) heat flow shot noise $S_H$, and (c) heat flow $H$, as functions of the Fermi energy. Their units are obtained by substituting $\hbar \omega =1$ meV into the energy and absorbing additional $2 \pi$ into the data. An inflection occurs at $E_F \approx 0.826$ meV corresponding to the resonance in transmission. Insets are the zoom-in of the inflection point. }
\end{figure}

\begin{figure}[h]
\includegraphics[height=10cm, width=14cm]{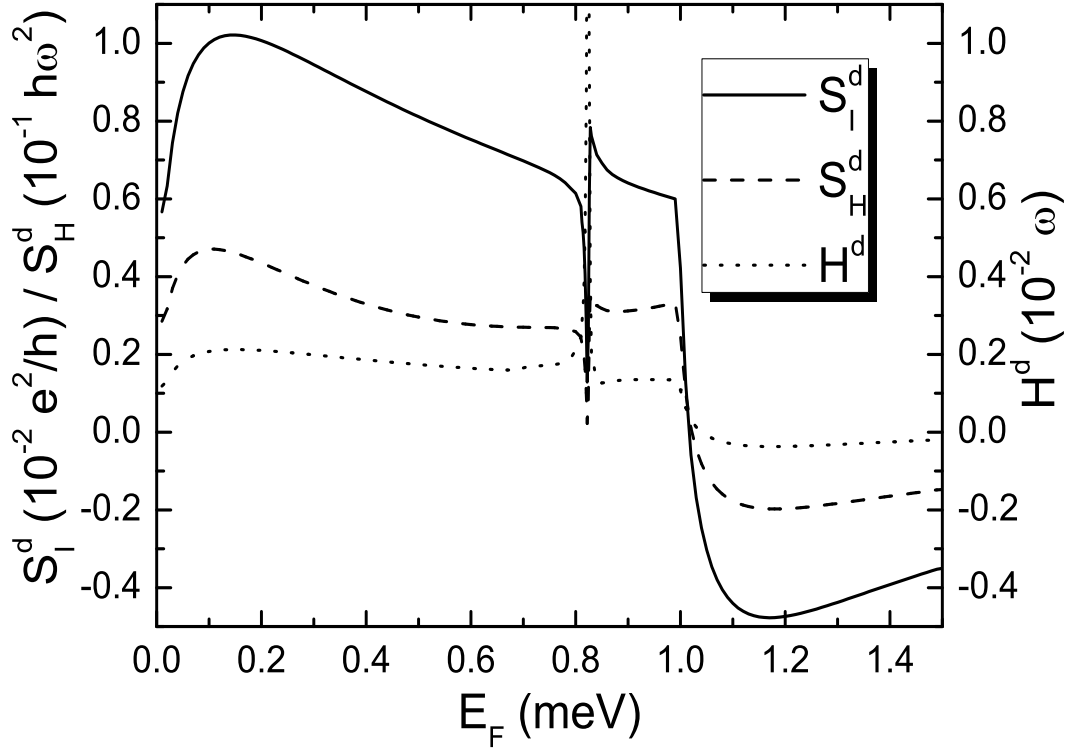}
\caption{Energy differentials of the current noise ${S^d_I}$, heat flow noise ${S^d_H}$, and the heat flow $H^d$, as functions of the Fermi energy. Their units are obtained by substituting $\hbar \omega =1$ meV into the energy and absorbing additional $2 \pi$ into the data. Sharp resonance could be seen at the inflection in Fig. 3 with $E_F \approx 0.826$ meV. }
\end{figure}

\clearpage

\end{document}